\documentclass[twocolumn,aps,pra]{revtex4}
\usepackage{epsfig}
\usepackage[english]{babel}
\usepackage{latexsym}
\usepackage{graphics}
\usepackage{subfigure}
\usepackage{graphicx}
\usepackage{dcolumn}
\usepackage{amsmath}
\usepackage{hyperref}
\usepackage{amssymb}
\usepackage{color}

\begin{document}

\title{Coulomb Effects on time-trajectory-resolved high-order harmonic generation}

\author{X. J. Xie$^{1}$, R. H. Xu$^{2}$, S. J. Yu$^{3,\dag}$,  X. Liu$^{4}$, W. Li$^{4}$, and Y. J. Chen$^{1,*}$}

\date{\today}
\begin{abstract}
We study the effect of Coulomb potential on high-order harmonic generation (HHG) numerically and analytically.
We focus on the influence of Coulomb potential on emission times of HHG associated with specific electron trajectories.
By using a numerical procedure based on numerical solution of time-dependent Schr\"{o}dinger equation (TDSE) in three dimensions,
we extract the HHG emission times both for long and short electron trajectories.
We compare TDSE predictions with those of a Coulomb-modified model arising from strong-field approximation (SFA).
We show that the Coulomb effect induces earlier HHG emission times than those predicted by the general SFA model without considering the Coulomb potential.
In particular, this effect influences differently on long and short electron trajectories and is more remarkable for low-energy harmonics than  high ones.
It also changes the HHG amplitudes for long and short electron trajectories.
We validate our discussions with diverse laser parameters and forms of Coulomb potential.
Our results strongly support a four-step model of HHG.
\end{abstract}

\affiliation{1.College of Physics and Information Technology, Shaan'xi Normal University, Xi'an 710119, China\\
2.Institute of Applied Physics and Computational Mathematics, Beijing 100094, China\\
3.School of Physics and Electronic Information, Yan'an University, Yan'an 716000, China\\
4.Beijing Institute of Space Mechanics and Electricity, Chinese Academy of Space Technology, Beijing, 100094, China}

\maketitle

\section{Introduction}

Owing to the advances of laser technology, the study of the nonlinear response of atoms and molecules to intense laser fields is a subject of much current interest
and significance in atomic, molecular, and optical physics \cite{1}.
When the interaction between the strength of laser field and electron-core is comparable, nonlinear processes could happen,
such as HHG \cite{2,3,4,5,6}, above-threshold ionization (ATI) \cite{7,8,9,10,11,12}
and non-sequential double ionization (NSDI) \cite{13,14,15}.
These processes have broad  applications in ultrafast measurements with attosecond resolution \cite{16,17}.
Particularly, the HHG can also be used to produce attosecond pulses, which open new prospect for probing the motion of electron insides atoms and molecules in its natural scale.
The HHG can be well understood through the classical three-step model (CM) \cite{4},
where harmonics are emitted through the following steps: tunneling ionization of the valence electron of the target,
propagation of the freed electron in the laser field and recombination of the freed electron with the parent ion in a rescattering event.
Another widely used approach to describe the HHG is SFA \cite{5}, which can be regarded as the quantum-version three-step model.

Both the CM and SFA with the saddle-point method describe the motion of the rescattering electron, responsible for HHG, in terms of electron trajectories.
These trajectories  include long and short trajectories related to the first return of the rescattering electron to the parent ion and multiple returns,
and are characterized by tunneling-out time and recombination time of the rescattering electron and harmonic energy. The recombination time of the rescattering electron with the parent ion also
corresponds to the emission time of a harmonic.
Based on these electron trajectories,  the attosecond dynamical information of the target can be deduced through HHG spectrum \cite{17,18}.
The CM and SFA neglect the effect of the Coulomb potential. However, a great deal of studies have shown that the Coulomb effect plays an important role
in the strong-field ionization process (i.e., the first step of HHG),  with remarkably affecting
photoelectron momentum \cite{19} and angular distributions \cite{20}, and
energy spectra \cite{21}.  In particular, by virtue of a semiclassical Coulomb-modified SFA model (MFSA),
recent studies on time-resolved strong-filed electron dynamics showed that the Coulomb effect can lead to
a significant lag of the ionization time of the system (about 100 attoseconds) relative to the peak time of
the laser electric field. This lag has important influences on HHG electron trajectories, with remarkably changing tunneling-out time and emission time.
Because of the basic importance of electron trajectories in the understanding of the HHG mechanism and the application of HHG in attosecond measurements,
detailed studies on the applicability of the MSFA predictions for HHG electron trajectories are highly desired.

In \cite{22}, with the numerical solution of  TDSE by spectral method \cite{23} in two-dimensional cases,
the short trajectory contributions are picked out and then the harmonic emission time of the short trajectory
is obtained through time-frequency analysis \cite{24,25}.
It is showed that, in comparison with SFA, the HHG emission time and amplitude of MSFA for short trajectory are in agreement with those of TDSE,
and the amplitude of  short trajectory predicted by MSFA is remarkably increased by about one-order magnitude \cite{22}.
In the following, we will call this numerical method that allows one to differentiate contributions of short trajectory
from those of long trajectory and multiple returns the short-trajectory TDSE simulation \cite{26}.
However,  the short-trajectory TDSE simulation are not capable of differentiating
the contributions of long trajectory from  multiple-return ones to HHG.
It therefore is unclear whether the HHG emission time and amplitude of  long trajectory  predicted by MSFA
are also consistent with the predictions of TDSE. In addition, in \cite{22}, it is only a small parameter region which is explored.

In this paper, with developing numerical procedures based on three-dimensional (3D) TDSE simulations,
which allow one to extract the HHG emission times and amplitudes of both short and long trajectories with high time resolution,
we are able to explore the effect of Coulomb potential on HHG electron trajectories and the applicability of MSFA in a wide parameter region.
Our 3D results show that, compared with SFA predictions, the emission times of  long and short trajectories predicted by MSFA  are closer to the TDSE ones.
In addition, the short-trajectory HHG amplitudes of MSFA  are also more consistent with  TDSE predictions than SFA in 3D cases,
 while the amplitudes for long trajectory predicted by all of MSFA, SFA and TDSE are near to each other.
In particular, with decreasing the range of the Coulomb potential in simulations, MSFA and TDSE predictions for short-trajectory amplitudes become near to
SFA ones, indicating that the form of Coulomb potential has important influence on HHG electron trajectory.
We show that this influence mainly arises from the fact that due to the existence of the Coulomb-induced ionization time lag which depends on the form of the Coulomb potential, the tunneling-out times of both short and long trajectories become earlier than the predictions of the general SFA and closer to the peak time of laser field on the whole. Instead of the general three-step model, our results strongly suggest a four-step model of HHG, including tunneling, lag, propagation and recombination.

\section{Theoretical methods}
\subsection{TDSE simulations }

\emph{Pseudospectral method}.-The TDSE is solved using the generalized pseudospectral method \cite{27}. We assume that the molecular axis is located in the xoy plane
and the laser field is linearly polarized along the direction parallel to the x axis. The Hamiltonian of the 3D  He atom studied here is
$\text{H}(t)={{{\text{\textbf{p}}}^{2}}}/{2}\;+\text{V}(\text{\textbf{r}})+\text{\textbf{r}}\cdot \text{\textbf{E}}(t)$ (in atomic units of $\hbar =e={{m}_{e}}=1$).
We use the soft-core potential that has the following form of $\text{V}(\text{\textbf{r}})=-{Z e^{-\rho r^{2}}}/{\sqrt{\xi +r^{2}}}$ with $r^{2}=x^{2}+y^{2}+z^{2}$.
$\rho$ is the screening parameter with $\rho=0$ for the long-range potential  and $\rho>0$ for the short-range one.
$\xi $=0.5 is the smoothing parameter which is used to avoid the Coulomb singularity and $Z$ is the effective charge
which is adjusted in such a manner that the ionization potential of the model system reproduced here is $I_p=0.9$ a.u..
$\mathbf{E}(t)= {{\vec{ \textbf{e}}}_x}f(t)E_{0}\sin{(\omega_0 t)}$ is the electric field. $f(t)$ is the envelope function,
$E_{0}$ and ${{\omega }_{0}}$ are the amplitude and the frequency of the laser field, and ${{\vec{\textbf{e}}}_{x}}$ is
the unit vector along the laser polarization which coincides with the $x$ axis. In our simulations, we use a trapezoidally
shaped laser pulse with a total duration of ten optical cycles and linear ramps of three optical cycles.
After each time step the TDSE wave function $\psi(\mathbf{r},t)$ of H(t) with $r \geq r_{0}$ is  multiplied by a mask function $ M(r)=\cos^{1/4}[\pi/2(r-r_{0})/(r_{m}-r_{0})]$ to absorb the continuum wave packet at the boundary. Here, $r_{0}$ is the boundary of the absorbing procedure and $r_{m}=400$ a.u. is the grid size \cite{28}.

The TDSE emission times of electron trajectories are obtained with finding the locally maximal amplitudes of time-energy distributions \cite{24,29}.  To obtain these distributions, as in \cite{30}, we project the TDSE wave function in the inner region (which is defined
as $\psi(\mathbf{r},t)$ with $r\leq 10$ a.u.) on the real-basis eigenstates of H$_{0}$ in each time step.
Relevant results are presented in Fig. 3 to Fig. 6 in the paper. For clarity,
we show only the simulations in which the electrons are emitted in the first half-optical cycle.

\emph{Spectral method}.-Alternatively, we can also solve the TDSE of $i\dot{\Psi}(t)=\text{H}(t)\Psi(t)$ numerically through the spectral method \cite{23}.
In this case, we work with a space grid size of $L_{x} \times L_{y} \times L_{z} = 102.4 \times 102.4 \times 51.2$ a.u. for the $x$, $y$, and $z$ axes, respectively.
The space step is 0.4 a.u. in all axes  and the time step is smaller than 0.1 a.u. in all wavelength cases. The mask function used here and the way
to obtain contributions of short trajectory are the same as introduced in \cite{31}.
Relevant results are presented in Fig. 7 in the paper.

\subsection{Analytical description}

\emph{MSFA}.-To analytically study the Coulomb effect on time-trajectory-resolved dynamics of HHG, we first calculate SFA-based ATI electron trajectories by the following saddle-point equations:
\begin{eqnarray}
{[\mathbf{p}+\mathbf{A}(t_s)]^2}/{2}=-I_p
\end{eqnarray}
with the complex  time $t_s=t_0+it_x$ and the drift momentum $\mathbf{p}$. Then we consider the Coulomb correction on ATI electron trajectories\cite{20,32}.  As in \cite{33,34}, we solve the Newton equation
\begin{eqnarray}
\ddot{\mathbf{r}}(\mathbf{p},t)=-\mathbf{E}(t)-\nabla_\mathbf{r} V(\mathbf{r})
\end{eqnarray}
for each SFA trajectory with initial conditions $\dot{\mathbf{r}}(\mathbf{p},t_0)=\mathbf{p}+\mathbf{A}(t_0)$ (the exit momentum) and $\mathbf{r}
(\mathbf{p},t_0)=Re(\int^{t_0}_{t_s}[\mathbf{p}+\mathbf{A}(t')]dt')$ (the exit position)  \cite{34, 35}, and the  instantaneous energy of the tunneling
electron as it travels is
\begin{eqnarray}
E_a(t)=[\dot{\mathbf{r}}(\mathbf{p},t)]^2/2+V(\mathbf{r}).
\end{eqnarray}
Here, $\mathbf{A}(t)$ is the vector potential of the laser field $\mathbf{E}(t)$, and $V(\mathbf{r})$ is the Coulomb potential as used in TDSE simulations.
The real part $t_0$ of the saddle-point time $t_s$  is considered as the tunneling-out time at which the electron exits the laser-Coulomb-formed barrier through
tunneling. The Newton equation Eq. (2) is solved using the Runge-Kutta method with adaptive step-size control. We will call the above Coulomb-modified SFA the MSFA.  In
our MSFA simulations  with $V(\mathbf{r})\neq0$  in Eq. (2), the ionization time $t_i$ is defined as the time when the value of  $E_a(t)$  becomes larger than zero for
the first time.   With assuming $V(\mathbf{r})\equiv0$, Eq. (2) can also be used to simulate the evolution of tunneling electrons related to the SFA electron trajectory
only in the laser field. In this SFA evolution, the  tunneling-out time $t_0$  agrees with the ionization time $t_i$ with the positive instantaneous energy $
[\dot{\mathbf{r}}(\mathbf{p},t_i=t_0)]^2/2$.  Because of the  presence of Coulomb potential, the MSFA predicts a time difference $t_{d}=t_i-t_0$ with $t_d>0$, which
influences remarkably on dynamics of the laser-driven system.  In contrast to SFA,  the ionization time of MSFA is slightly delayed \cite{22}.   We also obtain
Coulomb-modified HHG electron trajectories by finding the emission time $t_r$ which satisfies the relation $r(\mathbf{p},t_r)=0$ with $t_r>t_0$. The electron return
energy  and  amplitude at $t_r$ are $E_p=[\dot{\mathbf{r}}(\mathbf{p},t_r)]^2/2+V(\mathbf{r})$  and  $(1/\tau)^{1.5}e^{b}$ with $\tau=t_r-t_0$, respectively. Here, b is
the imaginary part of the ATI quasiclassical action $S(\mathbf{p},t_s)=\int_{t_s}^{t}dt'[(\mathbf{p}+\textbf{A}(t'))^{2}/2+I_{p}]$ with $t\rightarrow\infty$ at relevant
saddle points \cite{9} and only minus values of b are considered. This factor $(1/\tau)^{1.5}$ stands for the quantum-mechanical spreading of the wave
packet when it is traveling \cite{5}.

\emph{SFA}.-By comparison, in the general SFA model of HHG without considering the Coulomb effect, the electron trajectories are obtained with the following saddle-point equations
\begin{eqnarray}
{[\mathbf{p}_{st}(t_s,t_s')+\mathbf{A}(t_s)]^2}/{2}&=&-I_p,\\\nonumber
{[\mathbf{p}_{st}(t_s,t_s')+\mathbf{A}(t_s')]^2}/{2}&=&\Omega-I_p,
\end{eqnarray}
with  $\textbf{p}_{st}=-\int _{t_s}^{t_s'}dt'\textbf{A}(t')/(t_s'-t_s)$. The real parts of $t_s$ and $t_s'$ are considered as the SFA-based classical ionization time
 $t_i$ and emission time $t_r$ of the rescattering electron, respectively. The SFA amplitude for one harmonic with energy $\Omega$ is approximately evaluated with the
expression $(1/\tau)^{1.5}e^{b'}$. Here, $b'$ is the imaginary part of the HHG quasiclassical action
$S'(\textbf{p}_{st},t_s,t_s',\Omega)=\int_{t_s}^{t_s'}dt'[(\textbf{p}_{st}+\textbf{A}(t'))^2/2+\textbf{I}_{p}]-\Omega t_s'$
at relevant complex saddle points ($t_s,t_s'$) of Eq. (4) \cite{36} and only minus values of $b'$ are considered.
Note, the MSFA-based HHG amplitude $(1/\tau)^{1.5}e^{b}$ and the SFA-based one  $(1/\tau)^{1.5}e^{b'}$
are related to different actions of $S$ and  $S'$ and different saddle-point equations of Eq. (1) and Eq. (4), respectively.
In the following, for simplicity, the terms of ``TDSE" and ``MSFA" without especial illustration indicate simulations with the long-range Coulomb potential.

\begin{figure}[t]
\begin{center}
{\includegraphics[width=8.5cm,height=6.5cm]{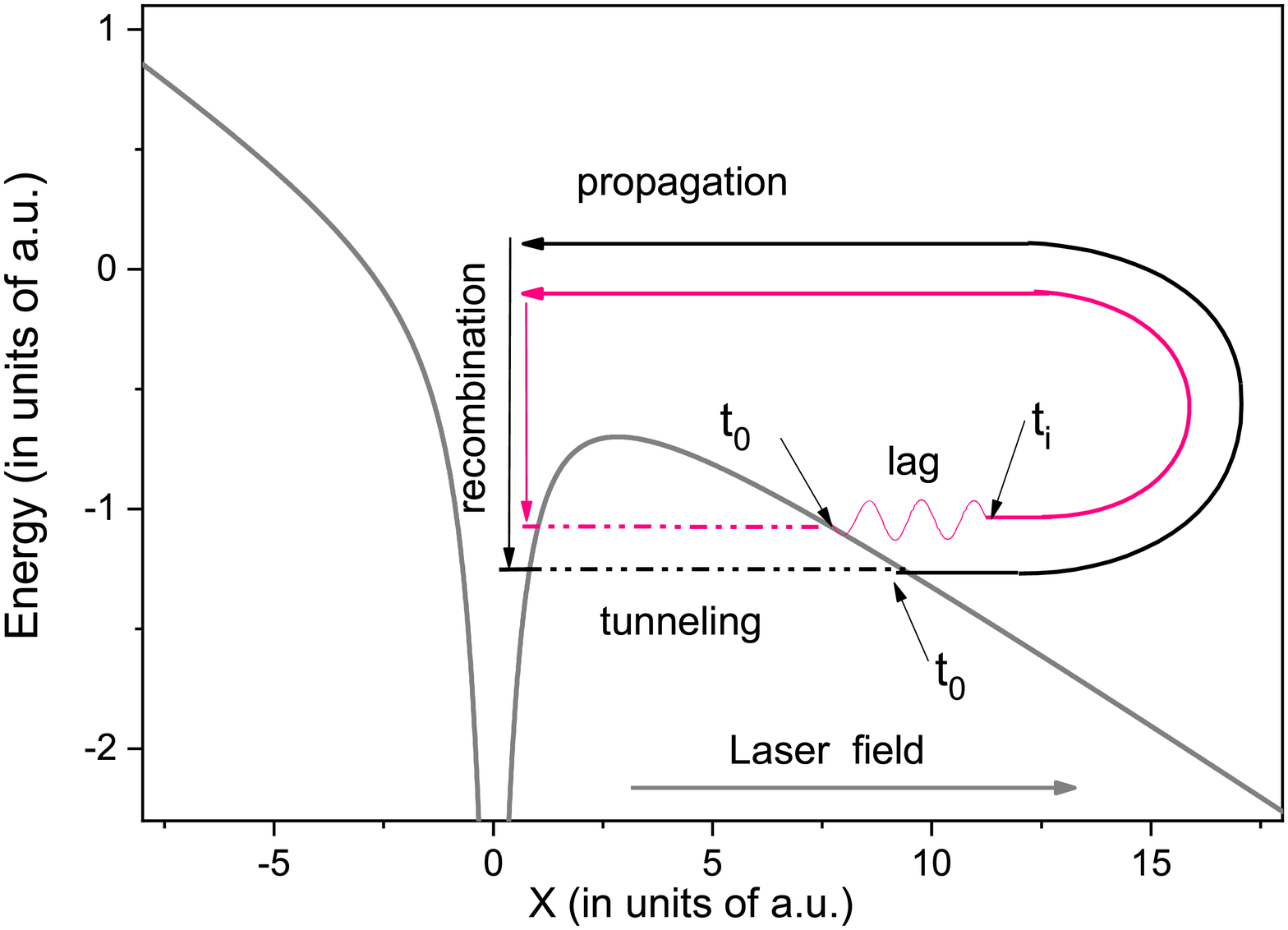}}
\caption {A sketch of HHG processes with and without Coulomb effect.  When the electron exits the laser-Coulomb-formed barrier (grey-solid curve) at a time $t_{0}$,  it can not be free immediately due to the Coulomb effect. Instead, it stays near the nuclei for a period of time $\bigtriangleup t$ (i.e., the lag of ionization time).  At the time $t_{i}=\bigtriangleup t+t_{0}$, the instantaneous energy of the electron becomes larger than zero and the electron moves far away from the nuclei. Then the electron is driven by the laser field to return to and recombine with the parent ions with the emission of a high-energy harmonic. Therefore, the HHG process can be understood as a four-step model (pink curve) including tunneling, lag, propagation and recombination instead of the well-known three-step model including tunneling, propagation and recombination, as indicated by the black curve.  It should be noted that the first step of HHG processes, i.e., tunneling, described by SFA (without Coulomb effect) and current MSFA (with Coulomb effect) are the same. Here, they are plotted by pink and black dashed-dotted lines  for clarity.  The laser field is along the direction parallel to the $x$ axis  as indicated by the gray solid arrow.}
\label{fig:graph1}
\end{center}
\end{figure}

\begin{figure}[t]
\begin{center}
{\includegraphics[width=8.5cm,height=5cm]{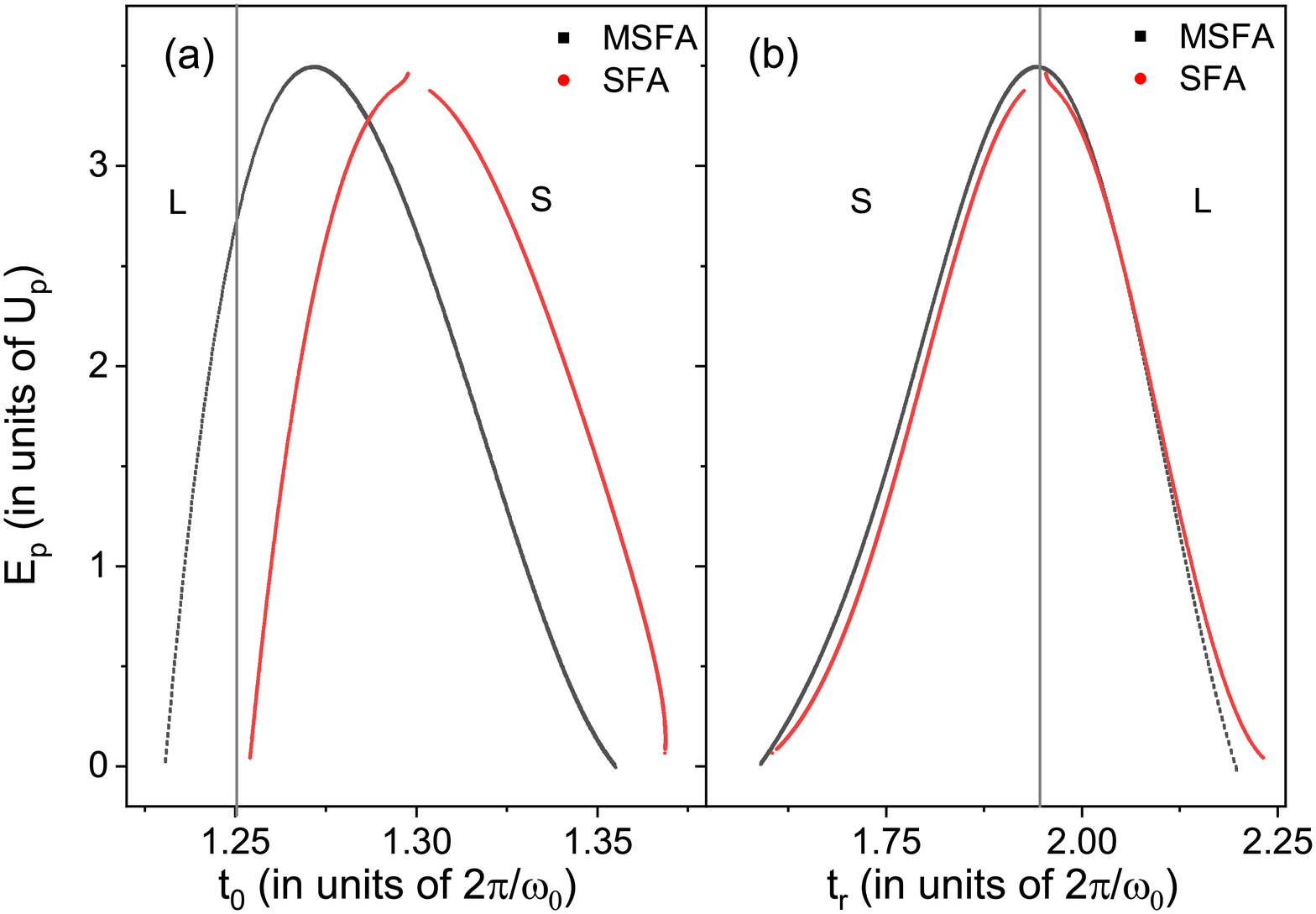}}
\caption{Effects of Coulomb-induced ionization time lag on HHG.  (a) The tunneling-out time $t_0$ versus the HHG return energy $E_p$ (scaled with $U_p=E_0^2/(4\omega^2_0)$) predicted by MSFA and SFA.  (b) The HHG return energy versus the emission time $t_r$ predicted by MSFA and SFA.   The vertical grey-solid line in (a) indicates the peak time of laser electric field. The one in (b) indicates the HHG cutoff positions, which divide the trajectories into long-trajectory (L) and short-trajectory (S) parts.  The laser parameters used are $I=5\times10^{14}$W/cm$^{2}$ and $\omega_0=0.057$ a.u..}
\label{fig:graph1}
\end{center}
\end{figure}

\begin{figure}[t]
\begin{center}
{\includegraphics[width=8.5cm,height=7cm]{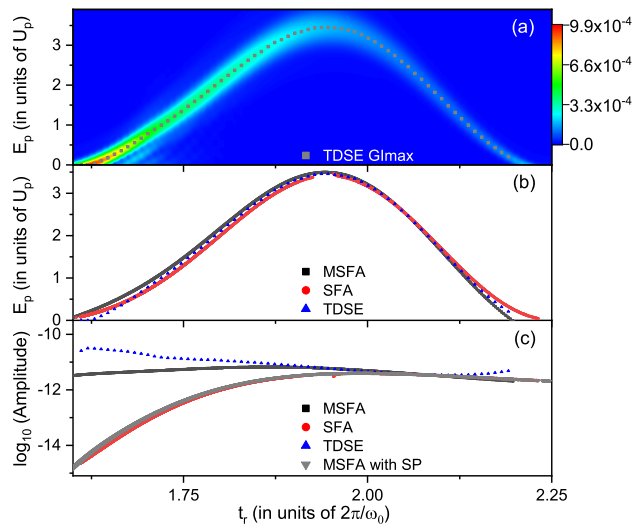}}
\caption {Comparisons of HHG long and short electron trajectories of He born at half a laser cycle, obtained with MSFA, SFA and TDSE.
Results  in (a) show the time-energy distribution of TDSE simulation. The gray square curve in (a) represents the global maximum amplitude of the distribution,
with which we obtain TDSE predictions of HHG emission time and amplitude.
These TDSE results are used as the reference to
compare the predictions of MSFA and SFA for HHG return energy $E_p$ (b) and amplitudes (c) versus the emission time $t_r$.
The gray curve in (c) indicates the results of MSFA with the short-range potential (SP). The TDSE curve in (c) is  shifted vertically for comparison. The laser parameters used are as in Fig. 2.}
\label{fig:graph1}
\end{center}
\end{figure}

\begin{figure}[t]
\begin{center}
{\includegraphics[width=8.5cm,height=7cm]{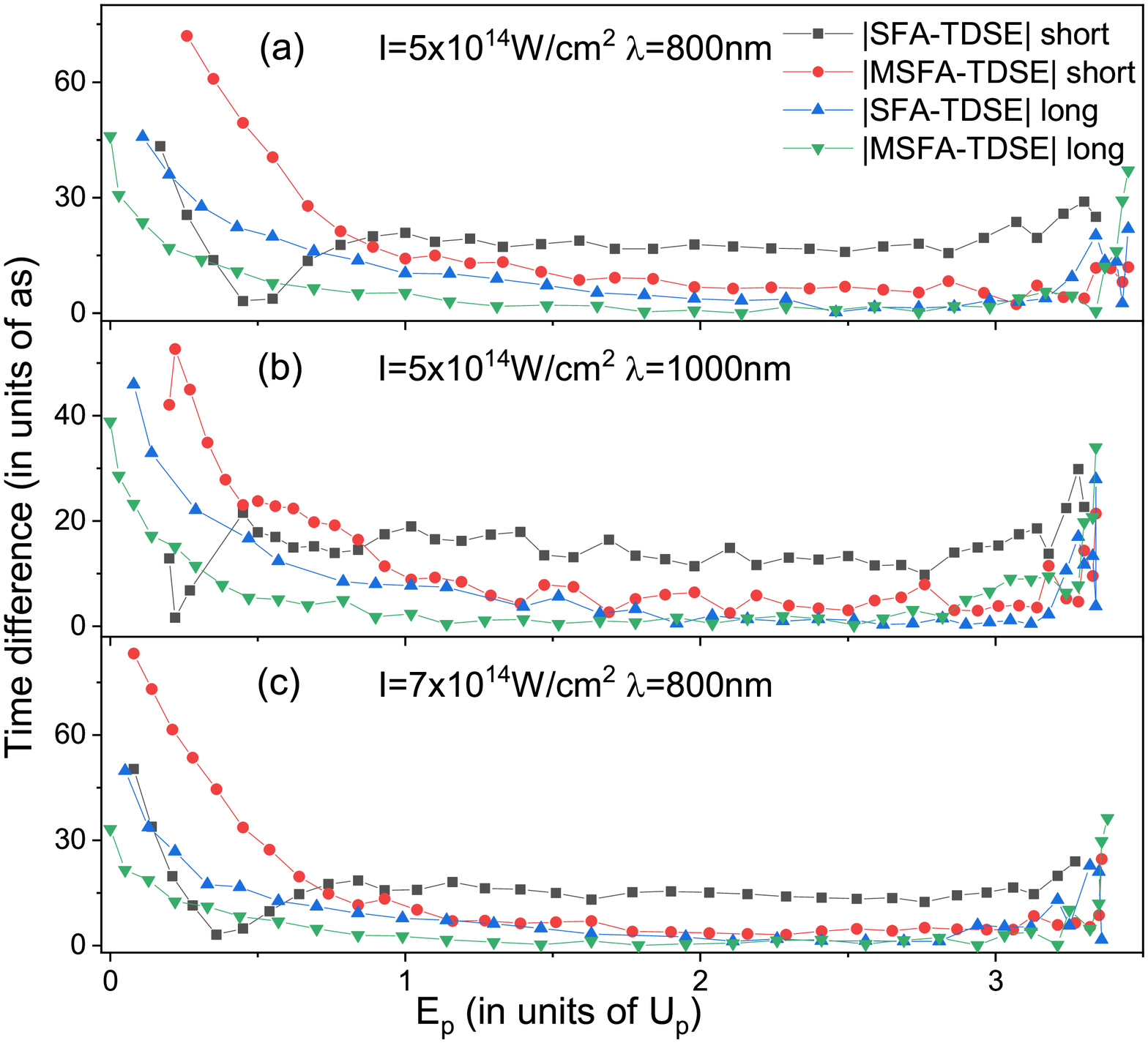}}
\caption{Comparisons of HHG return energy versus  emission time for short and long trajectories, calculated with MSFA, SFA and TDSE at different laser parameters. Here, we show the absolute value of difference between the results of TDSE and these models. The emission-time differences between SFA and TDSE of short and long trajectories are plotted by the black square curve and the blue up-triangle curve, respectively. Similarly, those between TDSE and MSFA  of short and long trajectories are represented by the red circle curve and the green down-triangle curve, respectively. The laser parameters used are as shown in each panel.}
\label{fig:graph1}
\end{center}
\end{figure}

\begin{figure}[t]
\begin{center}
{\includegraphics[width=8.5cm,height=7cm]{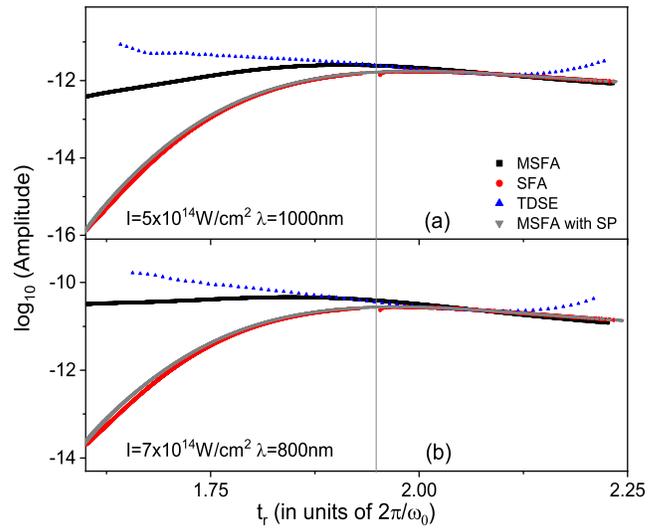}}
\caption{Same as Fig. 3(c) but for different laser parameters. The vertical grey-solid line  indicates the HHG cutoff positions, which divide the trajectories into
long-trajectory (right) and short-trajectory (left) parts. }
\label{fig:graph1}
\end{center}
\end{figure}

\begin{figure}[t]
\begin{center}
{\includegraphics[width=8.5cm,height=5cm]{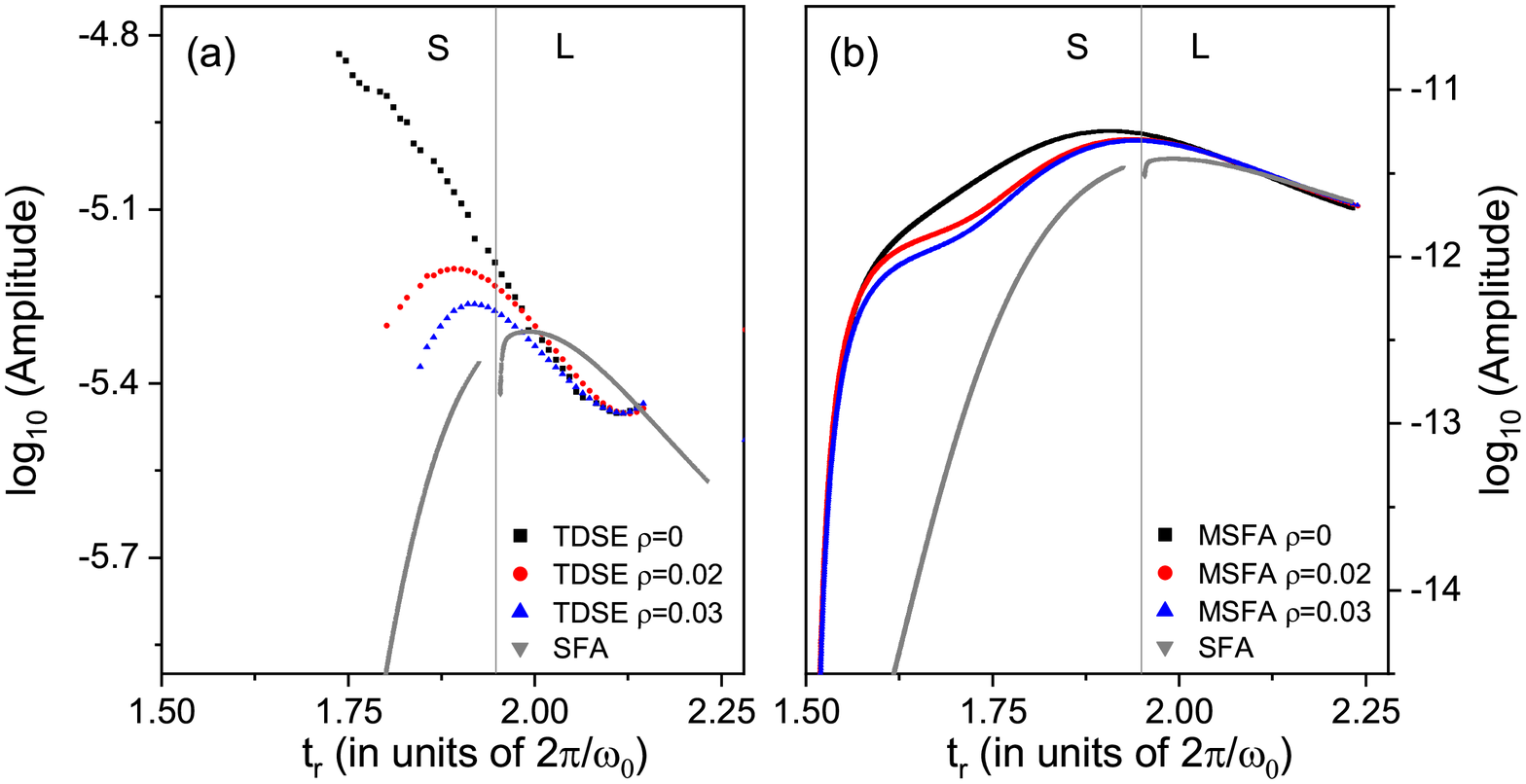}}
\caption{Comparisons for short- and long-trajectory HHG amplitudes versus emission time, calculated with TDSE (a) and MSFA (b) for different screening parameters $\rho$. For comparison, the SFA results (without Coulomb effect) are plotted in each panel. The vertical grey-solid lines indicate the HHG cutoff positions, which divide the trajectories into long-trajectory (L) and short-trajectory (S) parts. The laser parameters used are as in Fig. 2.}
\label{fig:graph1}
\end{center}
\end{figure}

\begin{figure}[t]
\begin{center}
{\includegraphics[width=8.5cm,height=7cm]{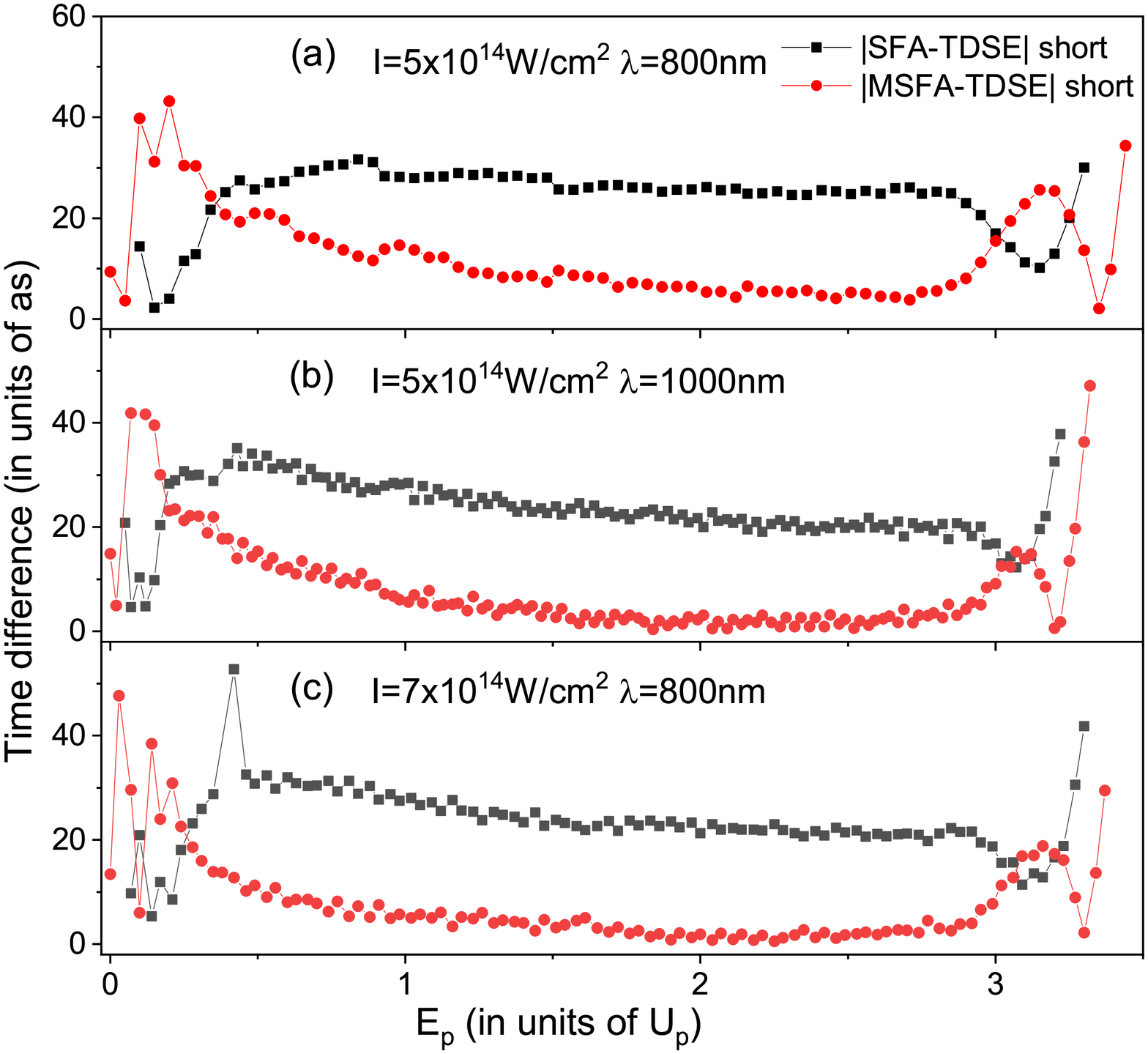}}
\caption{Same as Fig. 4, but  the results are obtained by TDSE solved numerically with spectral method in 3D and these results only include the contributions of short trajectory. }
\end{center}
\end{figure}

\section{Results and Discussions}

Figure 1 shows the sketch of HHG processes described by SFA and MSFA.
We assume that the laser field is along the $x$ axis.  According to the SFA where the Coulomb effect is neglected,
the HHG process includes these three steps of tunneling, propagation and recombination (black curve).
However, when taking the Coulomb effect into consideration, the Coulomb-induced ionization time lag also exists in the HHG process.
This lag means that the electron does not escape immediately after tunneling out of the laser-Coulomb-formed  barrier (grey-solid curve),
but vibrates near the parent ion for a period of time $\bigtriangleup t$ in the action of both the Coulomb field and the laser electric field,
and then escapes at a time $t_{i}=t_{0}+\bigtriangleup t$. Then the electron behaves similarly to a classical particle and is driven far away from the parent ion by the laser field.
When the laser field changes its direction, the electron can return to the vicinity of the nucleus and recombine with the parent ion with the emission of a high-energy photon.
The existence of the Coulomb-induced ionization time lag $\bigtriangleup t$ suggests the use of a four-step model (pink curve), including
tunneling, lag, propagation and recombination  to describe the HHG.

This lag discussed above has a profound influence on dynamics of the electron after tunneling in the strong laser field \cite{37,38}. As discussed in \cite{22} for 2D cases, this lag gives rise to the tunneling-out and emission times of HHG for both long and short electron trajectories dozens of attoseconds earlier than the predictions of SFA. In particular, the tunneling-out times of some long trajectories predicted by MSFA are earlier than the peak time of the laser field, while such trajectories disappear in SFA.  According to the SFA, the electrons which tunnel  out of the barrier at the rising part of the laser field
can not return and contribute to HHG.  However, when considering the influence of the Coulomb potential, these tunneling-out electrons can not be free immediately.  They stay near the parent nucleus for a period of time  and then escape at the falling part of laser field. So these electrons are also allowed to contribute to the rescattering event. Therefore, the tunneling-out times predicted by MSFA are advanced in comparison with SFA. The similar situation also goes for 3D simulations in the paper, as shown in Fig. 2, where we show the comparisons of  HHG electron trajectories obtained with SFA and MSFA of 3D Coulomb potential.
One can observe that in most energy regions in Fig. 2(a), the tunneling-out time predicted by MSFA in 3D cases is about 60 attoseconds earlier than that predicted by SFA.  At the same time, the emission time of MSFA  short trajectory is about 30 attoseconds  earlier than that of SFA in most energy regions and the emission-time difference between these two models for long trajectory is about 10 attoseconds, as shown in Fig. 2(b).  In 2D cases \cite{22}, the prediction of MSFA for the tunneling-out time of long or short trajectory is about 25 attoseconds earlier than the SFA one.  In addition, the emission time of HHG predicted by the MSFA for short trajectory is about 15 attoseconds earlier than the SFA one. These results suggest that the influence of Coulomb potential on  HHG electron trajectories  is  more remarkable in 3D cases  than in 2D ones and for short trajectory than for long one. When the tunneling-out time can not be easily to distill in TDSE simulations, the emission time is possible to access with time-frequency analyses, as discussed in the method part.

The 3D TDSE results are presented in Fig. 3.
In Fig. 3(a), we plot the time-energy distribution of 3D He atom following the procedure introduced in \cite{30}.
Here, the grey-square curve shows the global maximal amplitude of the distribution, and time and amplitude information related to this curve are defined as the TDSE emission time and amplitude of HHG. In Figs. 3(b) and 3(c), we compare these TDSE results to predictions of SFA and MSFA.
For results of return energy versus return time (emission time) in Fig. 3(b), the TDSE predictions are nearer to the MSFA ones on the whole.
Similar situations also hold for HHG amplitude versus return time (emission time) in Fig. 3(c). It is worth noting that  according to the predictions of MSFA, the  amplitudes of short and long trajectory are comparable to each other. In addition, the short-trajectory amplitudes of MSFA for return energy around 1$U_p$  are one order of magnitude higher than the SFA predictions.
The 3D results indicate that the Coulomb effect influences more remarkably on short trajectory than long one, similar to 2D cases.
For smaller return energy corresponding to harmonics near the threshold, the differences between TDSE and MSFA predictions are somewhat more remarkable.
It has been shown that near-threshold harmonics have complex origins beyond the description of quantum-trajectory theory \cite{39}.
For comparison, in Fig. 3(c), we also plot the predictions of MSFA with a short-range Coulomb potential. One can observe that
the short-range MSFA curve is very near to the SFA one, suggesting that our MSFA approach can return to SFA when the Coulomb effect is weak.

To highlight the difference between HHG emission time of MSFA and SFA, in Fig. 4, we show the comparison for different laser parameters, with using
TDSE emission time as a reference.
One can observe from Fig. 4(a) that for short trajectory, the emission-time difference between SFA versus TDSE (about 20 attoseconds) is larger than that
between MSFA versus TDSE (about 10 attoseconds) for electrons with energy higher than $1U_p$.
For long trajectory, the predictions of SFA, MSFA and TDSE are similar for higher energy near the HHG cutoff and differ somewhat from each other for lower energy.
The emission-time difference between SFA versus TDSE is about 5 attoseconds larger than that between MSFA versus TDSE for energy around 1$U_p$.
For energy near or lower than $0.5U_p$, the differences between TDSE and the models become remarkable due to the complex origin of near-threshold harmonics.
When increasing the laser wavelength or the laser intensity, the situation is similar, as shown in Figs. 4(b) and 4(c) with smaller keldysh parameters of $\gamma=\sqrt{I_p/(2U_p)}$. It seems that the differences between TDSE and these two models become smaller for smaller values of $\gamma$.

In Fig. 5, we further compare the HHG amplitudes of TDSE and these two models at other laser parameters corresponding to smaller Keldysh parameters.
Results presented here are similar to those in Fig. 3(c), with the predictions of MSFA nearer to the TDSE ones.

To identify roles of the form of Coulomb potential in HHG electron trajectories, in Fig. 6,
we plot comparisons between TDSE and MSFA with different screening parameters $\rho$.
The predictions of SFA are also plotted here as a reference.
One can  observe from Figs. 6(a) and 6(b), the amplitudes of TDSE and MSFA short trajectories
decrease as the values of $\rho$ increase and gradually approach the SFA predictions
when the amplitudes of TDSE and MSFA  long trajectories are similar to those of SFA,
implying a strong dependence of short trajectory amplitude on the screening parameter $\rho$.
Next, we give somewhat more discussions about the results of TDSE in Fig. 6(a).
Firstly, we only show TDSE amplitudes with smaller screening parameters, because the results of TDSE
in 3D depend strongly on the screening parameters and the HHG signals of large  screening parameters  are weak.
 Secondly,  the TDSE short-trajectory results
with low energy are absent. The possible reasons are as follows.  The ionization time of the short-trajectory electron with low energy  is related to lower laser intensity. At the same time, with short-range potential, the  ionization probability of the system is small and the signal of HHG is weak. In this case, there are other channels beyond SFA to  also contribute to HHG at lower energy, such as  channels associated with multiphoton ionization \cite{40,41,42,43}, in our TDSE simulations.  As a result, the contributions of  short trajectory at lower energy are difficult to pick out in TDSE simulations.   However, such channels are absent in MSFA simulations.

As a comparison, we also evaluate the HHG emission times obtained with short-trajectory TDSE simulations in 3D cases,
along with the TDSE solved by the spectral method.
Relevant results are presented in Fig. 7, where we also plot the differences between TDSE and these two models for different laser parameters.
The results are similar to those shown in Fig. 4.
One can observe from  Figs. 7(a) to 7(c), that the differences between TDSE and MSFA are remarkably smaller than those between TDSE and SFA and near zero for energy higher than 1$U_p$.
In addition, the differences of TDSE and MSFA seem smaller for cases of higher laser intensities or longer wavelengths corresponding to smaller values of $\gamma$.

\section{Conclusions}

In summary, we have studied the influence of the Coulomb potential on HHG electron trajectories for diverse laser parameters.
By using a numerical procedure based on 3D TDSE, we extract the HHG emission times both for long and short electron trajectories and compare
them to the predictions of the general SFA without considering the Coulomb potential and the MSFA where the Coulomb effect is included.
We have shown that the emission times of long (short) trajectory predicted by MSFA are about 10 (30) attoseconds earlier than those of SFA,
and the MSFA predictions are nearer to the TDSE ones. These results show that the Coulomb effect advances the emission time of both
long and short electron trajectories and  is more remarkable on short trajectory than long one.
In addition, the Coulomb effect is also more remarkable on low-energy harmonics than high ones, with the differences between predictions of
MSFA and SFA smaller for higher harmonic energy on the whole.
Our further analyses also show that, in comparison with SFA, the HHG amplitude of  MSFA  short trajectory is significantly increased,
while the HHG amplitude of MSFA long trajectory  is similar to SFA.
Moreover, the amplitude of the short trajectory  depends strongly on the form of the Coulomb potential.
These phenomena can be attributed to the Coulomb-induced ionization time lag, which advance the tunneling-out time of electron trajectories for both long and short ones
and make the tunneling-out time of short trajectory nearer to the peak time of the laser field.
The results provide insights into the mechanism of HHG and give suggestions on HHG-based attosecond measurements.

\section*{Acknowledgement}
National Natural Science Foundation of China (Grants No. 11904310, No. 91750111); the National Key Research and Development Program of China (2018YFB0504400).

\end{document}